\documentclass[12pt]{article}
\usepackage{cite}
\usepackage[dvips]{color}
\usepackage{epsfig}

\newcommand {\Was}{W\c as}
\newcommand {\KKMC}{\hbox{${\cal KK}$}\ MC}
\newcommand {\MISR}[1]{{\cal M}^{\rm ISR (#1)}_1}
\newcommand {\Sp}{{\rm Sp }}
\begin{document}
\begin{titlepage}
\begin{flushright}
{\bf BU-HEPP-04/07 }\\
{\bf October, 2004}\\
\end{flushright}

\begin{center}
{\Large
The Virtual Correction to Bremsstrahlung in High-Energy \lowercase{{\large e}}\raise0.8ex\hbox{\small+}\lowercase{{\large e}}\raise0.8ex\hbox{\small\bf--} Annihilation: Comparison of Exact Results
}
\end{center}

\vspace{2mm}

\begin{center}
{\bf S.A.\ Yost${}^a$, C.\ Glosser${}^b$, S.\ Jadach${}^c$ and 
B.F.L\ Ward${}^a$}
\end{center}

\vspace{2mm}
\begin{center}
{\em ${}^a$Department of Physics,
  Baylor University, Waco, Texas 76798-7316, USA,\\
 \em ${}^b$Department of Physics, Southern Illinois University, Edwardsville, IL 62026, USA\\
 \em ${}^c$Institute of Nuclear Physics,
        ul. Radzikowskiego 152, Krak\'ow, Poland,\\
 \em CERN, Theory Division, CH-1211 Geneva 23, Switzerland}
\end{center}
\vspace{5mm}
\begin{center}
{Presented by S.A. Yost at ICHEP 2004, International Conference on High Energy Physics, Beijing, August 16-22, 2004}
\end{center}
\vspace{5mm}
\begin{center}
{\bf Abstract}
\end{center}
\vspace{10mm}
We have compared the virtual corrections to single hard bremsstrahlung as
calculated by S.\ Jadach, M.\ Melles, B.F.L.\ Ward and S.A.\ Yost
to several other expressions.  The most recent of these comparisons is to
the leptonic tensor calculated by J.H.\ K\"uhn and G.\ Rodrigo for radiative
return.  Agreement is found to within $10^{-5}$ or better, as a fraction of
the Born cross section, for most of the range of photon energies. The massless
limits have been shown to agree analytically to NLL order.
\vspace{10mm}
\renewcommand{\baselinestretch}{0.1}
\footnoterule
\noindent
{\footnotesize
\begin{itemize}
\item[${\dagger}$]
Work partly supported
by the US Department of Energy Contract  DE-FG05-91ER40627 and by
NATO grant PST.CLG.980342.
\end{itemize}
}
\end{titlepage}

High precision studies of the Standard Model at proposed linear colliders
will require per-mil level control of both the theoretical and experimental
uncertainties in many critical processes to be measured. 
This will require computing higher order electroweak radiative
corrections at least to order ${\cal O}(\alpha^3L^3)$ for the
leading log effects and to the exact ${\cal O}(\alpha^2)$.
One important contributions at this order
is the virtual photon correction to the single hard bremsstrahlung
in $e^+e^-$ annihilations\cite{berends,in:1987,jmwy,hans1,hans2}. 

\begin{figure*}[ht]
\epsfig{file=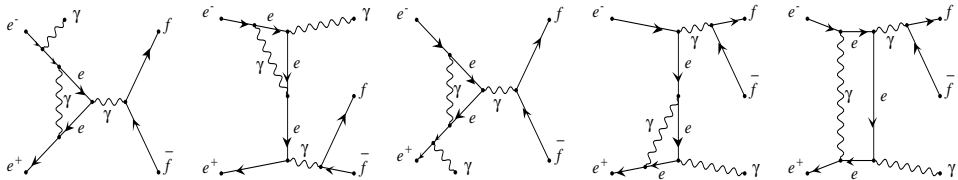,width=5in}
\caption{Feynman graphs for the virtual ${\cal O}(\alpha)$ correction to 
the process $e^+e^-\rightarrow f{\overline f}+\gamma$.}
\label{graph1}
\end{figure*}

In Refs.~\cite{jmwy}, some of us (S.J., B.F.L.W., S.A.Y.) have
presented comparisons of the results in Refs.~\cite{berends,in:1987,jmwy},
and in general a very good agreement was found. In particular, all of
these results can be shown to agree analytically to NLL order. However, 
some differences become apparent 
at the level of the NNLL (next-to-next-to leading log), 
depending on the different
levels of ``exactness'' in the calculations. Specifically,
the mass corrections are included in Ref.~\cite{jmwy} in a fully
differential way, whereas in Ref.~\cite{berends} the mass corrections
are included but the photon angular variable is integrated over and
in Ref.~\cite{in:1987} the results are fully differential
but the mass corrections are incomplete. These comparisons therefore
can not really test the NNLL, fully differential results
in Ref.~\cite{jmwy}.

More recently, another fully differential expression for the virtual photon
correction to single hard bremsstrahlung has appeared in 
Refs.~\cite{hans1,hans2}, where the result is obtained by an independent method
via a leptonic tensor calculated to investigate radiative return,
with particular emphasis on the 1-2 GeV cms energy regime.
This permits a cross-check at the NNLL level of the
corresponding results.  Details have been presented in 
Refs.~\cite{compare,paris}.  An indirect comparison of the
results in Refs.~\cite{jmwy,hans1,hans2} was also reported
in Ref.~\cite{hans3}  via a comparison of the two
Monte Carlo programs PHOKHARA~\cite{phokhara} and 
${\cal KK}$ MC~\cite{kkmc:2001}, as these two
Monte Carlos have the realizations of the 
results in Ref.~\cite{hans1,hans2} (PHOKHARA)
and Ref.~\cite{jmwy} ( ${\cal KK}$ MC). Agreement at the 
per-mil level was found on selected observables. 

Relevant Feynman graphs for the process $e^+ e^- \rightarrow f{\overline f}
+ \gamma$ are illustrated
in Fig.~\ref{graph1}.  In Ref.~\cite{jmwy} the ISR
matrix element is evaluated using the helicity spinor amplitude method of CALKUL\cite{berklei}, Xu {\it et al.}\cite{xuzhangchang} and Kleiss and Stirling\cite{KS}. Computer algebra techniques\cite{form} 
were used for reducing the loop integrals.

The ${\cal O}(\alpha^2)$ virtual correction to
single hard brem\-sstrah\-lung can be
expressed in terms of a form factor multiplying the ${\cal O}(\alpha)$
tree level matrix element\cite{jmwy}:
\begin{equation}
\label{eqn:misr}
\MISR{1} =  {\alpha\over 4\pi} (f_0 + f_1 I_1 + f_2 I_2)\MISR{0}
\end{equation}
where $\MISR{0}$ is the tree-level hard brem\-sstrah\-lung matrix element
and $\MISR{1}$ includes an additional virtual photon. The form factors 
$f_i$ and spinor factors $I_i$ can be found in Ref.~\cite{jmwy} or 
Ref.~\cite{paris}. (The latter form is equivalent to the earlier form,
but is explicitly numerically stable in collinear limits.) In the Monte
Carlo program, we actually calculate the YFS residuals\cite{yfs}
$\bar\beta_{1}^{(2)}$ which are obtained by subtracting the soft limit
$4\pi B_{\rm YFS}$ defined in Ref.~\cite{jmwy} from $f_0$ in Eq.~\ref{eqn:misr},
and using the matrix element to compute the cross-section.

The mass corrections are then added following the methods in
Ref.~\cite{berends1}, after checking that the exact expression
for the mass corrections differs from the result obtained by the
latter methods by terms which vanish as $m_e^2/s\rightarrow 0$
where $m_e$ is the electron mass and $s$ is the squared cms energy. 
The effect is to add a correction
\begin{eqnarray}
\label{eq:mass}
\langle f_0\rangle_{m} &=& {2m_e^2\over s} \left({r_1\over r_2} 
	+ {r_2\over r_1}\right) 
{z \over (1-r_1)^2 + (1-r_2)^2}
\nonumber\\
& &\times \left\{\langle f_0\rangle + (\ln z - 1) L
  - {3\over 2} \ln z + {1\over 2} \ln^2 z + 1\right\} 
\end{eqnarray}
to the spin-averaged form factor $\langle f\rangle_0$, where 
$L = \ln(s/m_e^2)$ is the ``big logarithm'' in the LL expansion, $z = s'/s$ and 
$r_i = 2p_i\cdot k/s$, with incoming fermion momenta $p_i$ and real photon
momentum $k$ defined as in Ref.~\cite{jmwy}. 

To NLL order, an 
expression for the cross section which correctly reproduces all collinear
limits can be obtained by setting $f_1 = f_2 = 0$ and using the spin-averaged
NLL form factor
\begin{eqnarray}
\label{NLL}
\langle f_0\rangle^{\rm NLL} &=& 2\left\{L - 1\right\}
+ {r_1(1-r_1)\over 1 + (1-r_1)^2}
+ {r_2(1-r_2)\over 1 + (1-r_2)^2}
+ 2\ln r_1 \ln (1-r_2)
\nonumber\\
&+& 2\ln r_2 \ln (1-r_1)
- \ln^2 (1-r_1)
- \ln^2 (1-r_2)
+ 3\ln (1-r_1)
\nonumber\\
&+& 3\ln (1-r_2)
+ 2\,\Sp(r_1) 
+ 2\, \Sp(r_2) + \langle f_0\rangle_m^{\rm NLL}
\end{eqnarray}
where $\Sp(x)$ is the Spence dilogarithm function and only terms surviving
when $r_1 \rightarrow 0$ or $r_2\rightarrow 0$ are needed in the mass 
correction. In the NLL limit, the ${\cal O}(\alpha^2)$ YFS residual
${\overline \beta}_1^{(2)}$ for real + virtual emission may be expressed in 
terms of the YFS residual ${\overline \beta_1}^{(1)}$ for pure real emission
using the result Eq.~\ref{NLL}, 
\begin{equation} {\overline \beta}_1^{(2)} = {\overline \beta}_1^{(1)} 
\left(1 + {\alpha\over 2\pi} \langle f_0\rangle^{\rm NLL}\right). 
\end{equation}

The new comparison uses the leptonic tensor of Ref.~\cite{hans2}, 
\begin{eqnarray}
\label{leptensor}
L_0^{\mu\nu} &=& {e^6\over s{s'}^2} \left\{ a_{00} s\eta^{\mu\nu} 
+ a_{11} p_1^\mu p_1^\nu 
+ a_{22} p_2^\mu p_2^\nu
+ a_{22} (p_1^\mu p_2^\nu + p_1^\nu p_2^\mu)\right.
\nonumber\\
 & &+\left. i\pi a_{-1} (p_1^\mu p_2^\nu + p_1^\nu p_2^\mu)\right\}
\end{eqnarray}
The coefficients $a_{ij}$ may be found in Ref.~\cite{hans2}
The squared matrix element for real photon emission can be obtained by 
contracting this with a final-state tensor
\begin{equation}
H^{\mu\nu} = e^2 (p_3^\mu p_4^\nu + p_3^\nu p_4^\mu
- p_3\cdot p_4 \eta^{\mu\nu})
\end{equation}
to obtain 
\begin{equation}
\label{kuhnCS}
|{\cal M}^{\rm ISR}|^2 = z L_0^{\mu\nu} H_{\mu\nu}.
\end{equation}

The coefficients $a_{ij}$  can be 
decomposed as $a_{ij} = a_{ij}^{(0)} + {\alpha\over\pi}a_{ij}^{(1)}$ in terms 
of a real-photon emission term and a virtual correction of order $\alpha$.
The  YFS residual ${\overline \beta}_1^{(2)}$ may be obtained by calculating
the cross section using Eq.~\ref{kuhnCS} with $a_{ij}^{(1)}$ replaced by 
$c_{ij} = a_{ij}^{(1)} - a_{ij}^{\rm IR}$, where the IR-divergent term is 
\begin{equation}
a_{ij}^{\rm IR} = a_{ij}^{(0)}\left[2(L-1)\ln v_{\rm min} + {1\over 2} L
- 1 + {\pi^2\over3}\right]
\end{equation}
with cutoff $v_{\rm min}$ on the fraction of the beam energy radiated into
the real photon, $v = 2E_\gamma/\sqrt{s}$. Again, it is possible to find
a relatively compact expression for the NLL contribution to the YFS residual
in the massless limit. The result is 
\begin{equation} {\overline \beta}_1^{(2)} = {\overline \beta}_1^{(1)} 
\left(1 + {\alpha\over 2\pi} C_1\right) + C_2 
\end{equation}
with coefficient functions
\begin{eqnarray}
C_1 &=& {1\over 2a_{00}^{(0)}}\left( {c_{11}\over z} + z c_{22} - 
	2c_{12}\right),
\nonumber\\
C_2 &=& {c_{11}\over 4z} + {zc_{22}\over 4} - {c_{12}\over 2} - c_{00},
\end{eqnarray}
where the IR-finite coefficients $c_{ij}$ are to be evaluated in the 
collinear limits $r_1 \rightarrow 0$ or $r_2 \rightarrow 0$. We have verified
that in these limits, $C_1 = \langle f_0\rangle^{\rm NLL}$ and $C_2 = 0$,
so that the results agree analytically in the massless NLL limit.

\begin{figure}[ht]
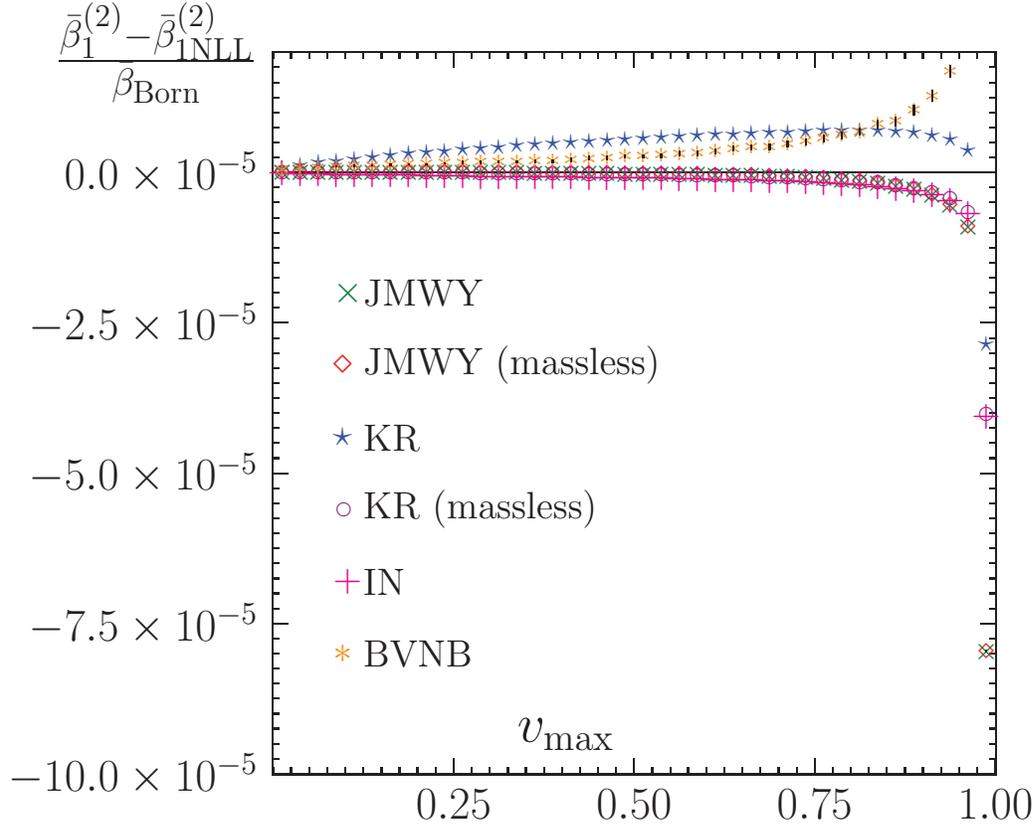

\include{4-0564-fig2}
\caption{
NNLL contribution
$\bar\beta_1^{(2)} -\bar\beta^{(2)}_{1 \rm NLL}$
for $10^8$ events generated by the YFS3ff Monte Carlo 
as a function of the cut $v_{\max}$ on the fraction
of the beam energy radiated to the photon. The results is
in units of the non-radiative Born cross section for 
$e^+ e^- \rightarrow \mu^+\mu^-$.
}
\label{fig:Fig2}
\end{figure}
The results are illustrated in Fig.~\ref{fig:Fig2} for the
case \hbox{$f\bar f = \mu^- \mu^+$}.
We show the 
complete $\bar\beta_1^{(2)}$ distribution
for our exact result JMWY as presented
in Ref.~\cite{jmwy}, both with and without mass corrections,
the result IN of Igarashi and Nakazawa {\it et al.}~\cite{in:1987}, 
the result BVNB of Berends {\it et al.}~\cite{berends}, and the new comparison
with exact results KR of K\"uhn and Rodrigo in Ref.~\cite{hans2} (with mass
corrections) and Ref.~\cite{hans1} (without mass corrections). 
The results were obtained using the YFS3ff MC generator 
(the EEX3 option of the \KKMC\ in 
Ref.~\protect\cite{kkmc:2001}) with $10^8$ events.
The NLL contribution calculated in Ref.~\cite{jmwy} has been subtracted
in each case to permit a clear NNLL comparison.
We see that there is a very good general agreement between all of 
these results.

Specifically, of the results agree to within
$0.4\times 10^{-5}$ for cuts below 0.75.  For cuts between 0.75 and .95,
the results agree to within $0.5\times 10^{-5}$, if the result
of Ref.~\cite{berends} is not included, and within $1.1\times 10^{-5}$ if
that result is included. For the last data point, at $v = 0.975$, the
result of Ref.~\cite{berends} is approximately $1\times 10^{-4}$ greater
than the others and is off-scale in Fig.~\ref{fig:Fig2}, while the 
remaining results agree to $3\times 10^{-5}$.
The difference between the KR and JMWY result attributable to differences
in the mass corrections never exceeds $10^{-5}$ over the entire range of
$v_{\max}$. 

These results are consistent with a total precision tag
of $1.5\times 10^{-5}$ for our ${\cal O}(\alpha^2)$
correction $\bar\beta_1^{(2)}$ for an energy cut below $v=0.95$. The NLL effect alone is adequate to within
$1.5\times 10^{-5}$ for cuts below 0.95.
The NLL effect has already been implemented in the
\KKMC\ in Ref.~\cite{kkmc:2001} and the attendant version of KK MC will
be available in the near future~\cite{jad2}.

These comparisons show that
we now have a firm handle on the precision tag
for an important part of the complete ${\cal O}(\alpha^2)$ corrections
to the $f{\overline f}$ production process needed for precision
studies of such processes in the final LEP2 data analysis, in the
radiative return at $\Phi$ and B-Factories, and in the
future TESLA/ILC physics.

\section*{Acknowledgments}
\label{acknowledgments}

Two of the authors (S.J.\ and B.F.L.W.) would like to thank 
Prof.\ G.\ Altarelli of the CERN TH Div.\ and Prof.\ D.\ Schlatter
and the ALEPH, DELPHI, L3 and OPAL Collaborations, respectively, 
for their support and hospitality while this work was completed. 
B.F.L.W.\ would like to thank Prof.\ C.\ Prescott of Group A at SLAC for his 
kind hospitality while this work was in its developmental stages. S.Y.\ would 
like to thank Prof.\ D.\ Bardin of the JINR, Dubna, for hospitality while
this work was completed.  Work partly supported 
by the US Department of Energy Contract DE-FG05-91ER40627 and by 
NATO grant PST.CLG.980342.

\label{references}

\end{document}